\documentclass[aps,preprint,showpacs,showkeys]{revtex4}
\usepackage{graphicx}
\usepackage{subcaption}
\usepackage{amsmath}
\usepackage{amsfonts}
\usepackage{amsthm}
\usepackage{mathrsfs}
\usepackage{amssymb}
\usepackage{graphicx}
\usepackage{epsfig}
\usepackage{hyperref}
\usepackage{color}
\usepackage{amscd}
\usepackage{xcolor}
\usepackage{fancyhdr}
\usepackage{xspace} 
\usepackage{float}
\usepackage{soul}
\usepackage{comment}

\begin{document}
\definecolor{myframecolour}{HTML}{FF0013}
\newcommand{\be}{\begin{equation}}
\newcommand{\ee}{\end{equation}}
\newcommand{\bea}{\begin{eqnarray}}
\newcommand{\eea}{\end{eqnarray}}

\def\beq#1{\begin{equation}\label{#1}}
\def\eeq{\end{equation}}

\newcommand{\bear}[1]{\begin{eqnarray}\label{#1}}
\newcommand{\ear}{\end{eqnarray}}

\newcommand{\R}{{\mathbb R}}
\newcommand{\p}{\partial}
\newcommand{\nn}{\nonumber}

\title{Shadow, Emission, and Strong-Field Lensing of Dilatonic Black Holes} 

\author{Kuantay Boshkayev$^{1,2}$, Ainur Urazalina$^{1}$, \\Aidana Kurmanbek$^1$, Manas Khassanov$^{1}$, Daniya Utepova$^{3}$ and \\ Yassine Sekhmani$^{4,5,6}$
}

\affiliation
{ $^1$ Institute for Experimental and Theoretical Physics, \\ Al-Farabi Kazakh National University, Almaty, 050040, Kazakhstan \\ }

\affiliation
{$^2$Kazakh‐-British Technical University, Tole bi str.~59, Almaty, 050000, Kazakhstan\\
}

\affiliation
{$^3$ Abai Kazakh National Pedagogical University, Almaty, 050010, Kazakhstan \\
}

\affiliation{$^4$Fesenkov Astrophysical Institute, Observatory 23, 050020 Almaty, Kazakhstan.\\}
\affiliation{$^5$Center for Theoretical Physics, Khazar University, 41 Mehseti Street, Baku, AZ1096, Azerbaijan.\\}
\affiliation{$^6$Centre for Research Impact \& Outcome, Chitkara University Institute of Engineering and Technology, Chitkara University, Rajpura, 140401, Punjab, India}

\begin{abstract}

We study the shadow and strong-field optical properties of a static, spherically symmetric dyon-like dilatonic black hole. The photon sphere radius, critical impact parameter, and shadow radius are obtained and analyzed in terms of the charge $Q$ and the dilatonic coupling parameter $a$. We show that increasing these parameters decreases the photon sphere and shadow radii, leading to a smaller apparent shadow. The predicted angular diameter is compared with the observational data for M87$^{*}$ and SgrA$^{*}$. We explicitly show that the Sgr A* data do not constrain the model parameters, whereas the results for M87* depend on the adopted mass estimates from stellar and gas dynamics. We also estimate the high-frequency energy emission rate in the geometric-optics approximation and derive the leading Bozza coefficient $\bar{a}$, which characterizes the logarithmic behavior of the deflection angle in the strong-field regime. Astrophysical implication of the obtained outcomes are discussed.

\end{abstract}

\maketitle

\section{Introduction}
\label{sec:intro}
Black-hole shadows provide a useful way to study photon motion in the strong-field regime. The Event Horizon Telescope observations of M87$^{*}$ and SgrA$^{*}$ have made it possible to compare theoretical shadow models with observational data and to test possible deviations from the standard black-hole solutions of general relativity \cite{EventHorizonTelescope:2019dse,EventHorizonTelescope:2022wkp}. The boundary of the shadow is directly related to unstable null geodesics and to the critical impact parameter of photons. Therefore, the photon sphere and the shadow radii are important quantities for studying both classical and non-standard black-hole spacetimes \cite{Luminet1979,bardeen1973four,cunha2018shadows,Perlick:2021aok}.

In modified and extended theories of gravity, black holes may possess additional parameters associated with scalar fields, vector fields, nonlinear electrodynamics, or quantum-gravity-inspired corrections \cite{bambi2017black,cardoso2019testing}. One important class of such objects is represented by dilatonic black holes. These solutions appear in gravitational models containing scalar fields coupled to electromagnetic or Abelian gauge fields through dilatonic coupling vectors. The corresponding spacetime geometry may differ from the Schwarzschild or  Reissner--Nordstr\"om geometries, while still reducing to them in appropriate limiting cases. The role of dilatonic fields and color charges in such black-hole solutions has been studied in detail in numerous works (see e.g. \cite{BronShikin,GibW,GM,GHS,ChHsuL,BBFM,BI,PTW,Br0,IMp3,LY,Ifbb,ABDI,ABI,AIMT,GalZad}).

The dyon-like dilatonic black hole considered in this work is characterized by the gravitational mass $M$, the net charge $Q$, and the dilatonic coupling parameter $a$. Depending on the value of $a$, this family of solutions contains several important limiting cases. For $a=0$, the metric reduces to the Schwarzschild solution; for $a=1$, it is related to the static Sen black hole; and for $a=2$, it reduces to the Reissner--Nordstr\"om metric after an appropriate coordinate transformation \cite{malybayev2021quasinormal,boshkayev2024circular,BTIMNU,IKMN,IMNT,2025JPhCS3089a2012B,1992PhRvL..69.1006S,2025EPJC...85.1477B,2020JPhCS1690a2143B}. This makes the model useful for comparing the optical properties of classical and dilatonic black holes within a single parametrized framework.

Previous studies of these dilatonic black holes have mainly focused on circular geodesics, innermost stable circular orbit radii, stability conditions, quasinormal modes, and numerical lensing images \cite{2020JPhCS1690a2143B,malybayev2021quasinormal,boshkayev2024circular,Beisenbekova_Urazalina_Khassanov_Utepova_Quevedo_2026}.
It has been shown that the parameters $Q$ and $a$ can noticeably modify the photon sphere and the structure of null geodesics. Since the shadow radius is determined by the critical impact parameter, these modifications can directly affect the apparent size of the black-hole shadow. This motivates a more detailed analysis of the shadow geometry and related optical characteristics of dilatonic black holes.

The main purpose of the present work is to investigate the observational and optical properties of this class of dyon-like dilatonic black holes, with particular emphasis on black-hole shadows, emission properties, and strong gravitational lensing. The previous studies provide the theoretical basis for our analysis; however, the present manuscript extends this framework in several directions.

More specifically, the main contributions of the present work are as follows:

1) We introduce a representation of the line element in terms of the parameters $M$ and $Q$, together with the coupling parameter $a$, which is more convenient for an observational analysis. In the original parametrization, the solution is expressed in terms of the parameters $\mu$ and $P$. These parameters are useful for constructing the exact solution, but they are not directly related to the quantities that are typically inferred from astrophysical observations. By rewriting the metric in terms of $M$ and $Q$, we obtain a parametrization that can be more directly connected with observationally relevant black-hole properties.

2) We calculate the black-hole shadow radius for the considered spacetime and investigate its dependence on the model parameters. This provides the basis for a direct comparison between the theoretical predictions of the model and black-hole shadow observations.

3) We investigate the possibility of constraining the model parameters, in particular the coupling parameter $a$ and the charge parameter $Q$, using the available Event Horizon Telescope observations of the supermassive black-hole candidates M87$^*$ and Sgr A$^*$.

4) We calculate the black-hole emission rate and study how the corresponding emission properties depend on the parameters of the spacetime. This provides an additional observable characteristic of the considered black-hole geometry.

5) We investigate the strong-field gravitational lensing properties by calculating the Bozza coefficients and analyzing the corresponding lensing observables. To the best of our knowledge, this provides an additional optical characterization of the considered dyon-like dilatonic black-hole spacetime that complements the previous studies of photon spheres and circular null geodesics.

Thus, we fully acknowledge that the underlying black-hole solution and some of its basic geodesic properties are not new. At the same time, the present work focuses on a different set of observables and, in particular, on their possible connection with astrophysical observations, including black-hole shadows, emission properties, and strong-field lensing.

The paper is structured as follows. In Sect.~\ref{sec:spacetime}, we review the main characteristics of dyon-like dilatonic black-hole spacetimes, discuss their key parameters, and introduce a new representation of the solution by expressing the metric functions in terms of the total mass $M$ and net charge $Q$. In Sect.~\ref{sec:phsh}, we derive the photon-sphere radius and black-hole shadow radius as functions of $M$, $Q$, and $a$. In Sect.~\ref{sec:obs}, we use observational data from the Event Horizon Telescope to constrain the model parameters. In Sect.~\ref{sec:emis}, we investigate the energy emission rate and black-hole thermodynamics. In Sect.~\ref{sec:Bozz}, we analyze the light-deflection angle in the strong-field regime and calculate the Bozza coefficients. Finally, in Sect.~\ref{sec:con}, we summarize our conclusions and discuss the astrophysical implications of our findings.

\section{Dyon-like black hole spacetime}
\label{sec:spacetime}
In this section, we consider a static and spherically symmetric dyon-like black hole spacetime discussed in Refs.~\cite{2020JPhCS1690a2143B,malybayev2021quasinormal}. 

The action of a model with two scalar fields, 2-form and 
dilatonic coupling vectors is given by 
\bear{i.1}
 S= \frac{1}{16 \pi G}  \int d^4 x \sqrt{|g|}\biggl\{ {\cal R}[g] -
  g^{\mu \nu} \p_{\mu} \vec{\varphi}  \p_{\nu} \vec{\varphi}
 \qquad \qquad   \nn \\
 - \frac{1}{2} e^{2 \vec{\lambda}_1 \vec{\varphi}} F^{(1)}_{\mu \nu} F^{(1)\mu \nu }
 - \frac{1}{2} e^{2 \vec{\lambda}_2 \vec{\varphi}} F^{(2)}_{\mu \nu} F^{(2) \mu \nu}
 \biggr\},
\ear
where $G$ is the gravitational constant, $g= g_{\mu \nu}(x)dx^{\mu} \otimes dx^{\nu}$ is the metric,  $|g| =   |\det (g_{\mu \nu})|$ is the determinant of the metric tensor, 
 $\vec{\varphi} =  (\varphi^1,\varphi^2)$ is the vector of scalar fields 
 belonging to ${\R}^2$, 
 $F^{(i)} = dA^{(i)}  =  \frac{1}{2} F^{(i)}_{\mu \nu} dx^{\mu} \wedge dx^{\nu}$
is the $2$-form with $A^{(i)} = A^{(i)}_{\mu} dx^{\mu}$, $i =1,2$, 
$\vec{\lambda}_1 = (\lambda_{1i}) \neq \vec{0}$, $\vec{\lambda}_2 = (\lambda_{2i}) \neq \vec{0}$ are the dilatonic coupling  vectors satisfying the following condition 
 \beq{i.1a}
 \vec{\lambda_1} \neq - \vec{\lambda_2} 
 \eeq
and {\bf ${\cal R}[g]$} is the Ricci scalar ~\cite{2020JPhCS1690a2143B}. Here the speed of light in vacuum is set $c =1$.

We focus on a  dyonic-like black hole
solution of the  field equations corresponding to the action
(\ref{i.1}) \cite{2020JPhCS1690a2143B}
\bear{i4.9}
 ds^{2} &=&
\mathcal{H}^{a}
\left[
-\mathcal{H}^{-2a}
\left(1-\frac{2\mu}{r}\right)dt^2
+
\frac{dr^2}{1-\frac{2\mu}{r}}
+
r^{2}\left(d\theta^2+\sin^2\theta d\varphi^2\right)
\right],
 \\ \label{eq:ds}
 \varphi^i &=& \nu^i \ln \mathcal{H} , 
\ear
\vspace{-0.5cm}
\beq{i4.10em}
 F^{(1)} = \frac{Q_1}{\mathcal{H}^2 r^2} dt \wedge dr, \qquad
  F^{(2)}  =  Q_2 \tau,    
\eeq
where $Q_1$ and $Q_2$ are (color) charges - electric and magnetic, respectively,
 $\mu > 0$ is the extremality parameter,
 $\tau = \sin \theta d \theta \wedge d \phi$ is the standard volume form on $S^2$,  
   \beq{i4.7}
     \mathcal{H} = 1 + \frac{P}{r},
   \eeq
   is the moduli function, with $P > 0$ fulfilling the equations
   \beq{i4.8a}
     P (P + 2 \mu) = \frac{1}{2} Q^2,
   \eeq
   \beq{i5.1}
 GM =   \mu + \frac{a}{2} P.
\eeq
Hereafter we set $G=1$.
All the remaining parameters of the solution are defined by
\bear{i4.10a}
   a &=&  \frac{  ( \vec{\lambda}_1 + \vec{\lambda}_2 )^2}{ \Delta },
 \\  \label{i4.10n}
     \Delta &\equiv& \frac{1}{2}  (\vec{\lambda}_1 + \vec{\lambda}_2)^2 +
      \vec{\lambda}_1^2 \vec{\lambda}_2^2 - (\vec{\lambda}_1 \vec{\lambda}_2)^2,
\\ \label{i2.BB} 
\nu^i &=& \frac{ \lambda_{1i} \vec{\lambda}_2 ( \vec{\lambda}_1 + \vec{\lambda}_2 )
    - \lambda_{2i} \vec{\lambda}_1 ( \vec{\lambda}_1 + \vec{\lambda}_2 )}{ \Delta }, 
  \ear
$i = 1,2$ and
 \beq{i4.8c}
    Q_1^2  = \frac{ \vec{\lambda}_{2}( \vec{\lambda}_1 + \vec{\lambda}_2 )}{2 \Delta} Q^2, \qquad
    Q_2^2  = \frac{ \vec{\lambda}_{1}( \vec{\lambda}_1 + \vec{\lambda}_2 )}{2 \Delta} Q^2.
 \eeq

As one may notice, the line element is described by three parameters $\mu$, $P$ and $a$ which encode all the information about the spacetime geometry \cite{2020JPhCS1690a2143B}. At the same time this form of the line is generic for dilatonic dyon black holes and dyon-like black holes in the model with two Abelian
gauge fields and two scalar fields. Therefore, we mainly focus on the general form of the line element. 

One can explicitly show that using Eqs.~\eqref{i4.8a}-\eqref{i5.1}  the parameters $\mu$ and $P$ are related to the gravitational mass $M$ and the net charge $Q$ of the black hole as follows
\begin{subequations}
\begin{align}
\mu&=\frac{M}{1-a}\left[1-\frac{a}{2}\left(1+\sqrt{1+\frac{(1-a)Q^2}{2M^2}}\right)\right].
\label{eq:Param}\\
P&=-\frac{M}{1-a}\left(1-\sqrt{1+\frac{(1-a)Q^2}{2M^2}}\right),
\label{eq:mu}
\end{align}
\end{subequations}
For the purposes of the present work, it is more convenient to use the parametrization in terms of $M$ and $Q$, since these quantities are directly relevant for the analysis of the photon sphere, shadow radius, and strong-field lensing coefficients. 
Here we consider both non-extremal and extremal black hole cases, which obey $Q^2/M^2 < 8/a^2$ and $Q^2/M^2 = 8/a^2$  conditions, respectively, for $0 < a \leq 2$ \cite{malybayev2021quasinormal}.

Introducing the compact notation $\Delta=\sqrt{2M^{2}+(1-a)Q^{2}},$ the metric can be rewritten in the standard form
\begin{equation}
\label{eq:ds_2}
ds^{2}=
-F(r)dt^{2}
+
\frac{dr^{2}}{F(r)}
+
H(r)\left(d\theta^{2}+\sin^{2}\theta d\phi^{2}\right).
\end{equation}
Here the metric functions are given by
\begin{equation}
\label{Eq:FGH_compact}
F(r)=
\frac{1-\frac{r_2}{r}}
{\left(1-\frac{r_1}{r}\right)^a},
\qquad
H(r)=
r^2\left(1-\frac{r_1}{r}\right)^a ,
\end{equation}
with
\begin{equation}
\label{Eq:r1r2_def}
r_1=
\frac{\sqrt{2}M-\Delta}{\sqrt{2}(1-a)} < 0,
\qquad
r_2=
\frac{\sqrt{2}(2-a)M-a\Delta}{\sqrt{2}(1-a)} >0.
\end{equation}
This compact representation will be used throughout the paper in the analysis of null geodesics, the shadow radius, the emission rate, and the strong-field lensing coefficients.
Several well-known black hole solutions are readily recovered as particular limits of this metric. For ($a=0$), the spacetime reduces to the Schwarzschild solution. For ($a=1$), the corresponding limiting form is equivalent to the Sen black hole solution~\cite{1992PhRvL..69.1006S,2025EPJC...85.1477B}. In the limiting case when $a\to1$, the quantities $r_1\to - Q^2/(4M)$ and $r_2\to2M-Q^2/(4M)$. For ($a=2$), after the coordinate transformation ($r=r_{\rm RN}-P$), the metric reduces to the Reissner-Nordström form
\begin{equation}
\label{eq:ds_3}
ds^{2}=-\left(1-\frac{2M}{r}+\frac{Q^{2}}{2r^{2}}\right)dt^{2}
+
\left(1-\frac{2M}{r}+\frac{Q^{2}}{2r^{2}}\right)^{-1}dr^{2}
+
r^{2}\left(d\theta^{2}+\sin^{2}\theta d\phi^{2}\right).
\end{equation}
Thus, the geometry considered here provides a unified family of black hole spacetimes interpolating between the Schwarzschild, Sen, and Reissner--Nordström cases. The relation between the parameter $a$, the dilaton coupling, and the interpretation of the charge $Q$ in terms of color charges is discussed in detail in Refs.~\cite{2020JPhCS1690a2143B,malybayev2021quasinormal,boshkayev2024circular}.

For the present metric, the event horizon is determined by the condition $F(r_{h})=0$. Since $F(r)$ contains the factor $1 -r_{2}/r$, one obtains $r_{h}=r_{2}$, or explicitly

\begin{equation}\label{Eq:r_h}
r_{h}=r_{2}=\frac{\sqrt{2}(2-a)M-a\Delta}{\sqrt{2}(1-a)} .
\end{equation}

\begin{figure}[htbp]
    \centering
  \includegraphics[width=0.45\textwidth]{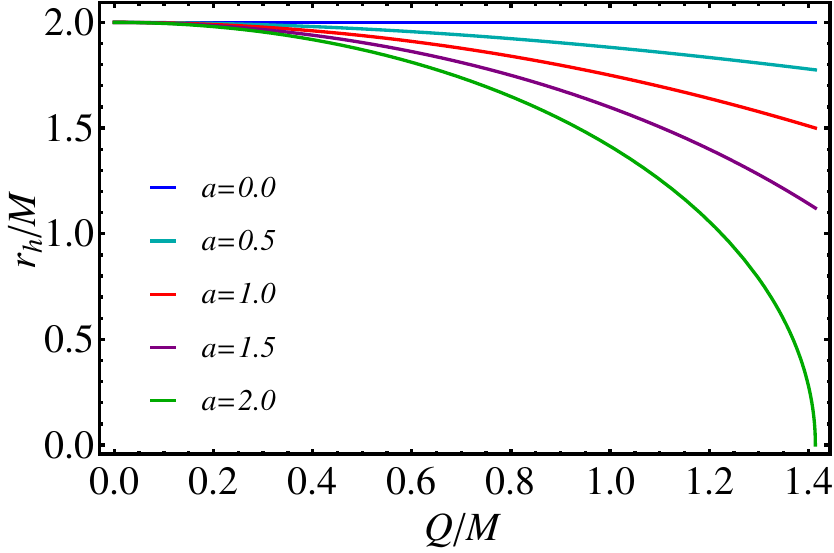}
    \caption{The plot illustrates the horizon radius as functions of the parameters $Q$ and $a$.}
    \label{fig:rh}
\end{figure}

In Fig.~\ref{fig:rh} we show the horizon radius $r_h/M$ of a dilatonic BH versus $Q/M$ for different values of the parameter $a$. 
As one can see, in the Schwarzschild case corresponding to $a=0$, the horizon radius is given by $r_h=2M$ and is independent of the charge. In the extremal Reissner--Nordstr\"{o}m case, when $a=2$ and $Q/M=\sqrt{2}$, the horizon radius becomes $r_h=0$, in agreement with the known results.

\section{The Photon Sphere and Black Hole Shadow}
\label{sec:phsh}

We now investigate the shadow properties of dilatonic BHs. The BH shadow, recently observed by the EHT \cite{EventHorizonTelescope:2019dse}, represents one of the most direct observational signatures of strong gravitational fields. For a distant observer, it appears as a dark silhouette whose boundary is determined by the photon region—the set of unstable null trajectories separating photons that fall into the BH from those that escape to infinity. In spherically symmetric spacetimes, such as the charged dilatonic BH considered here, the photon region reduces to a photon sphere \cite{Perlick:2021aok,Khodadi:2022pqh}, whose projection onto the observer's sky defines the shadow contour. Our goal is to determine how the dilatonic coupling parameter $a$ and the charge $Q$ affect the shadow geometry, particularly its radius $r_{\text{sh}}$ \cite{bernardo2023dressed}.

To characterize null motion, we confine our analysis to the equatorial plane ($\theta = \pi/2$) without loss of generality, which allows us to identify two conserved quantities for a particle: the energy $E = F(r)\dot{t}$ and the angular momentum $L = H(r)\dot{\phi}$.
For null geodesics ($ds^2=0$), the radial motion is expressed as:
\begin{equation}
\left(\frac{dr}{d\phi}\right)^2 = \frac{H(r)}{G(r)}\left(\frac{h(r)^2}{b^2}-1\right),
\end{equation}
where $G(r)=1/F(r)$, $h(r)^2 = H(r)/F(r)$ and $b=L/E$ is the impact parameter \cite{Perlick:2021aok}.

The photon sphere radius $r_{\text{ps}}$ is determined by the condition $h'(r)=0$, which yields the following quadratic equation:
\begin{equation}
 k_2r^2+k_1r+k_0=0
\end{equation}
where
\begin{eqnarray}
    k_0&=&(3-2a)\left((1-a)aQ^2+4M^2-2\sqrt{2}M\Delta\right)\, , \nonumber\\
    k_1&=&(1-a)\left(\sqrt{2}(2+a)\Delta-2(8-5a)M\right) \, ,\nonumber\\
    k_2&=&4(1-a)^2 \, .
\end{eqnarray}
Hence the quadratic equation will have two roots
\begin{equation}\label{rps}
r_{\text{ps}}=\frac{1}{2k_2} \left(-k_1\pm\sqrt{k_1^2-4k_2k_0}\right).
\end{equation}
We take the root with positive sign as it reduces to $3M$ for vanishing $a$ and $Q$.

It is worth noting that at first sight it seems that $r_{\rm ps}$ diverges in the Sen metric limit $a\to1$ as a direct substitution may cause ambiguity and confusion. This can be easily solved by finding the limit

\begin{equation}
\lim_{a\to 1}r_{ps}=\frac{24 M^2-3 Q^2+\sqrt{576 M^4-80 M^2 Q^2+Q^4}}{16 M}
\end{equation}

In all our computation for the case $a=1$, we used analogous limits to avoid indeterminate forms.

For a static observer at distance $r_o$, the apparent angular radius $\alpha_{\text{sh}}$ of the shadow is defined by:
\begin{equation}\label{angle2}
\sin^{2}\alpha_{\text{sh}}=\frac{b_{c}^{2}F(r_{o})}{H(r_{o})},
\end{equation}
where $b_c$ is the critical impact parameter. The shadow radius is then:
\begin{equation}
R_{\text{sh}}(r_o)=r_o\sin\alpha_{\text{sh}}
=r_o \sqrt{\frac{H(r_{\rm ps})F(r_o)}{F(r_{\rm ps})H(r_o)}}.
\end{equation}

For an observer at infinity ($r_o\to\infty$), this simplifies to:
\begin{equation}\label{eq:rsh}
R_{\text{sh}}(\infty)=r_{\text{sh}} = b_{cr}=\sqrt{\frac{H(r_{\rm ps})}{F(r_{\rm ps})}}.
\end{equation}

This expression is formally equivalent to the photon critical impact parameter $r_{\rm{sh}}=b_c$.

\begin{figure}[htbp]
    \centering
     \includegraphics[width=0.45\textwidth]{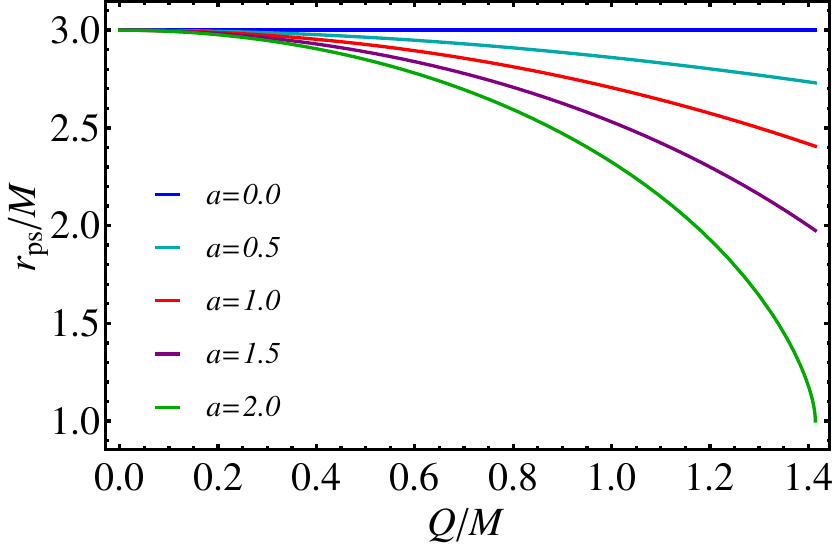}
    \vspace{0.5cm}
    \includegraphics[width=0.45\textwidth]{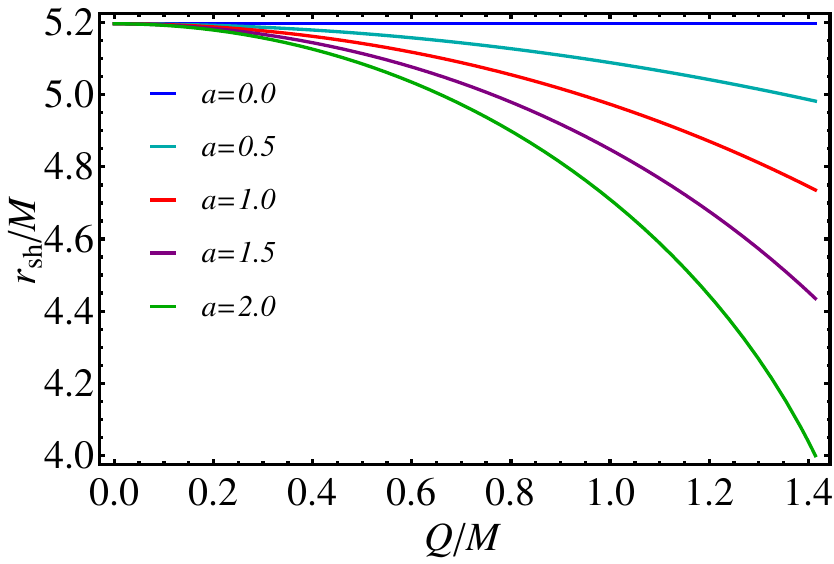}
    \vspace{0.5cm}
    \caption{The plot illustrates the variation of the photon sphere radii and shadow radius (critical impact parameter) as functions of the black hole charge $Q$ for different values of the dilatonic coupling constant $a$.}
    \label{fig:rph_rsh}
\end{figure}

Figure \ref{fig:rph_rsh} illustrates how the photon sphere radius $r_{\text{ps}}$ (left panel)  and shadow radius $r_{\text{sh}}$ (right panel) vary with respect to $Q$ for different values of $a$. Both radii decrease monotonically with increasing $Q$ and $a$. Consequently, larger values of the charge parameter lead to a smaller black-hole shadow, producing more compact circular silhouettes. Physically, the presence of electric charge, in addition to mass, modifies the spacetime geometry and alters the effective gravitational field experienced by photons. As a result, the unstable circular photon orbit shifts closer to the black hole, leading to a reduction in both the photon-sphere and shadow radii.

\begin{figure}[htbp]
    \centering   \includegraphics[width=0.45\textwidth]{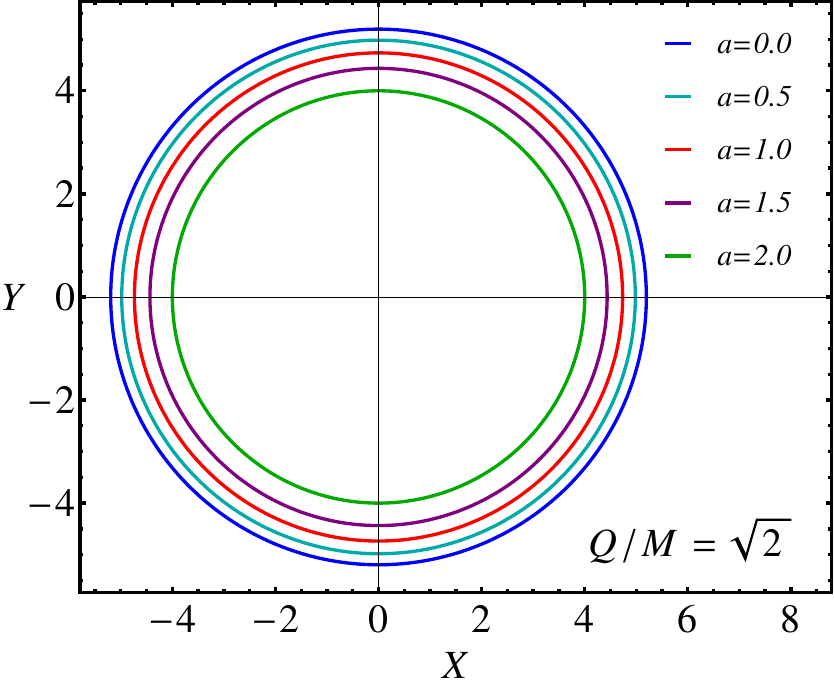}
    \vspace{0.5cm}
    \caption{Shadow silhouettes of the BH as seen by a distant observer for different values of the coupling parameter $a$. Moreover, the outer black photon sphere is associated with the Schwarzschild spacetime.}
    \label{fig:phsph}
\end{figure}

Fig.~\ref{fig:phsph} provides a visual representation of the shadow silhouettes for fixed $Q$ and different values of $a$. 

As $a$ increases, the shadow size decreases noticeably. This visual representation clearly demonstrates how the dilatonic parameter modifies the apparent size of the BH as seen by distant observers.

\section{Observations from EHT Data}
\label{sec:obs}
To connect the theoretical predictions with observational evidence, the following section presents an analysis of the shadow cast by dilatonic black holes, incorporating results from the Event Horizon Telescope collaboration. In particular, we use the EHT data for M87$^*$ and Sgr A$^*$ to constrain the parameters of these black holes. The apparent size of the black hole shadow, as measured by a distant observer, is characterized by its angular diameter \cite{2022PhR...947....1P}:
\begin{equation}
\left( \frac{\Omega^{*}}{\mu\text{as}} \right) = \left( \frac{6.191165 \times 10^{-8}}{\pi} \frac{M/M_\odot}{D/\text{Mpc}} \right) \left( \frac{b_{\text{c}}}{M} \right),
\label{eq:omega_relation}
\end{equation}
where $D$ is the BH's distance from the observer and $b_{\text{c}}$ is the critical impact parameter defined in Eq.~\eqref{eq:rsh}. Using shadow measurements from the EHT, it is possible to find constraints on the dilatonic parameters $a$ and $Q/M$. 

\begin{figure*}[t]
\begin{minipage}{0.49\linewidth}
\center{\includegraphics[width=0.97\linewidth]{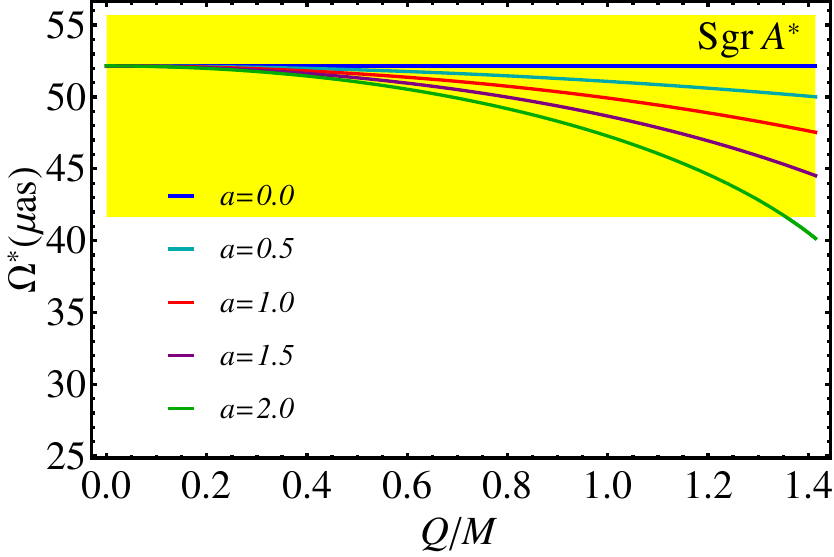}}\\ 
\end{minipage}
\hfill 
\begin{minipage}{0.50\linewidth}
\center{\includegraphics[width=0.97\linewidth]{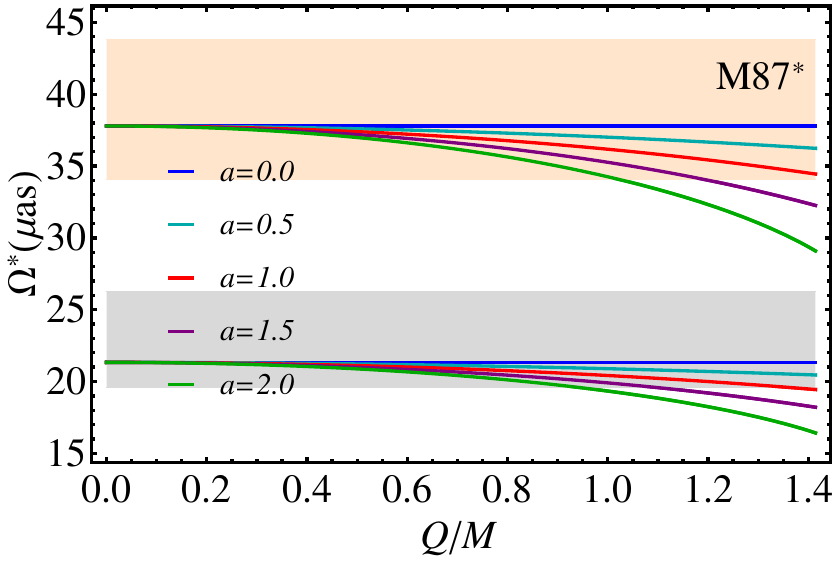}}\\ 
\end{minipage}

\caption{The relationship between the coupling parameter $a$, the BH charge $Q/M$, and the observed shadow's angular diameter for SgrA$^{*}$ (left panel), and M87$^{*}$ (right panel).}\label{fig:M87}
\end{figure*}

Figure \ref{fig:M87} (left panel) shows the constraints obtained from the Sgr A* observations. The yellow region corresponds to the measured shadow diameter, $48.7 \pm 7.0$ $\mu$as \cite{EventHorizonTelescope:2022wkp,2020PhRvD.101d1301B}. Using the mass estimate $M=4.0^{+1.1}_{-0.6}\times10^6 M_\odot$ and distance $D=8.127$ kpc for Sgr A* \cite{EventHorizonTelescope:2019dse}, we find that all considered values of $a$ and $Q/M$ within the range $0<Q/M<\sqrt{2}$ are consistent with the observational constraints. Therefore, the Sgr A* data do not provide strong restrictions on these model parameters. The analysis shown in Fig.~4 is similar to that performed in \cite{2021PhRvD.103j4047K,2023CQGra..40p5007V}, though here we focus exclusively on the black hole shadow size.

For M87*, the comparison requires additional care because the measured angular diameter of $42\pm3$ $\mu$as refers to the emission ring rather than directly to the black hole shadow. The EHT-derived mass, $M=(6.5\pm0.7)\times10^9 M_\odot$, including systematic uncertainties, and the distance, $D=(16.8\pm0.8)$ Mpc, imply a Schwarzschild shadow diameter of approximately $39.5\pm4.2$ $\mu$as \cite{EventHorizonTelescope:2019dse}. This estimate provides a useful reference for comparison with the predictions of our model, although the emission-ring and shadow diameters are not identical observables.

Independent mass measurements yield different estimates of the expected Schwarzschild shadow diameter. Using the stellar-dynamical mass, $M=6.2^{+1.1}_{-0.6}\times10^9 M_\odot$ (which is close to the EHT-derived mass) \cite{EventHorizonTelescope:2019dse}, we obtain $37.6^{+6.2}_{-3.5}$ $\mu$as, represented by the light-orange region in the right panel of Fig.~\ref{fig:M87}. In contrast, the gas-dynamical mass, $M=3.5^{+0.9}_{-0.3}\times10^9 M_\odot$ \cite{EventHorizonTelescope:2019dse}, corresponds to a shadow diameter of $21.3^{+5.0}_{-1.7}$ $\mu$as for a fixed distance to M87*, represented by the light-gray region. These differences illustrate the dependence of the inferred shadow size on the adopted mass estimate.

The uncertainties associated with the mass measurements lead to substantial allowed ranges of shadow diameters, thereby limiting the strength of the constraints on the model parameters. Nevertheless, the comparison indicates that shadow observations of M87* can, in principle, constrain the parameters $a$ and $Q/M$, although the strength of these constraints depends on the adopted mass estimate.

It is worth noting that, from a theoretical standpoint, for different values of $a$, the ratio $Q/M$ may exceed unity without violating physical consistency conditions (see, for further details, \cite{boshkayev2024circular}). These results emphasize the capability of black hole shadow observations to test dilatonic gravity and place bounds on its parameters. Moreover, the parameter-constraint analysis illustrated here should not be interpreted as a rigorous statistical parameter estimation. Our intention is to show the sensitivity of the predicted shadow size to $a$ and $Q/M$, rather than to perform a full Bayesian inference.

The observational constraints obtained here are complementary to our results in the subsequent section on light deflection and strong lensing, offering a coherent picture of how dilatonic gravity modifies the optical properties of black holes across different observational regimes \cite{wei2020extended}.

\section{The energy emission rate}
\label{sec:emis}
In this section, we examine the energy emission properties of dilatonic black holes. The shadow of a black hole provides valuable information about the high-frequency absorption behavior of the underlying spacetime, as the shadow area constitutes an excellent approximation to the absorption cross section of massless fields in the geometric-optics regime. In particular, for spherically symmetric black holes, the total absorption cross section \(\sigma(\omega)\) oscillates around a constant asymptotic value \(\sigma_{\rm lim}\) when the frequency greatly exceeds the inverse gravitational radius. Therefore, in the leading geometric-optics approximation, one may identify
\begin{equation}
\sigma_{\rm lim}\simeq\pi r_{\rm sh}^2,
\end{equation}
where \(r_{\rm sh}\) represents the shadow radius (equivalently, the critical impact parameter \(b_c\)) measured by a distant observer \cite{Wei:2013kza}. This approximation ignores low-frequency greybody effects and the intricate interference structure arising from partial waves; nevertheless, it accurately describes the leading contribution to the absorption cross section in the high-frequency regime. Employing \(\sigma_{\rm lim}\) , the spectral energy emission rate of a BH for a single bosonic degree of freedom can be approximated by a Planck-like expression, where the geometric cross section serves as an effective radiating area:
\begin{equation}\label{Eq:emission}
\frac{d^{2}E(\omega)}{dt\,d\omega}
\;=\;\frac{2\pi^{3}\,\omega^{3}\,r_{\rm sh}^{2}}{e^{\omega/T_H}-1}\,,
\end{equation}
where \(\omega\) denotes the frequency and \(T_H\) is the Hawking temperature. For completeness, the Hawking temperature may be expressed in terms of the surface gravity at the event horizon as \(T_H=\kappa/(2\pi)=F'(r_h)/4\pi\) (in units where \(G=\hbar=c=k_B=1\)), indicating that the characteristics of the model enter through both \(r_{\rm sh}\) and \(T_H\). Here, \(r_h\) denotes the horizon radius.

The Hawking temperature is obtained from the surface gravity at the horizon and can be written as
\begin{equation}
T_H=
\frac{F'(r_h)}{4\pi}
=
\frac{1}{4\pi r_h}
\left(1-\frac{r_1}{r_h}\right)^{-a}.
\end{equation}

\begin{figure}[htbp]
    \centering
    \includegraphics[width=0.45\textwidth]{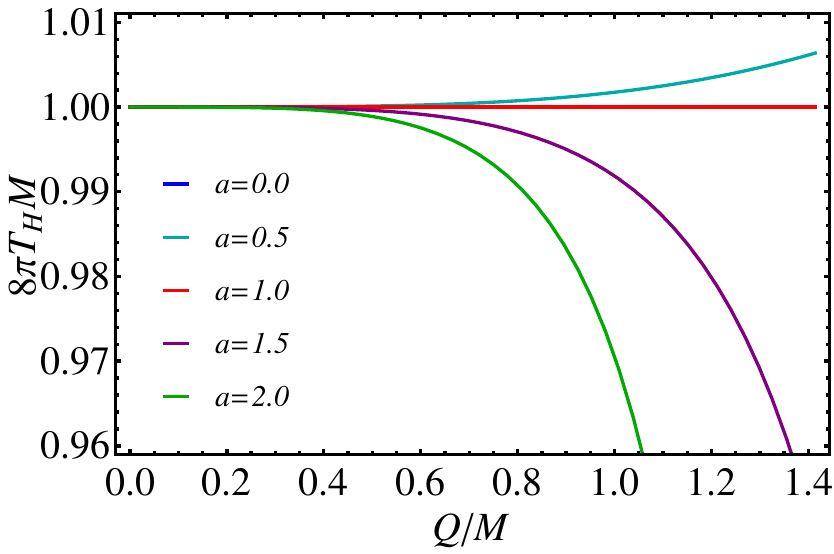}
    \caption{The Hawking temperature versus $Q$ for different values of $a$.}
    \label{fig:TH}
\end{figure}

Therefore, the dimensionless quantity used in Fig.~\ref{fig:TH}  is
\begin{equation}
\label{Eq:dimensionless_TH}
8\pi M T_H
=
\frac{2M}{r_h}
\left(1-\frac{r_1}{r_h}\right)^{-a}.
\end{equation}

In the Schwarzschild limit $Q=0$, this expression gives $r_{h}=2M$ and $T_{H}=1/(8\pi M)$, as expected. It is easy to show that in the limit $a\to1$ the Hawking temperature for the Sen metric equals the one of the Schwarzschild metric.

Finally, when a more accurate quantitative description is needed (for instance, to compare spectral profiles or peak frequencies across different \((a,Q)\) configurations), the geometric cross section should be replaced by the complete frequency-dependent absorption cross section \(\sigma(\omega)\), with the corresponding greybody transmission factors taken into account. Nevertheless, Eq.~\eqref{Eq:emission} provides the most straightforward and physically transparent approximation to the high-frequency emission spectrum.

In Fig.~\ref{fig:TH} we present the Hawking temperature $8\pi T_HM$ of a dilatonic black hole as a function of $Q/M$ for various values of the parameter $a$. One can notice a degeneracy between the cases $a=0$ and $a=1$. Moreover, for $a=0.5$ the temperature increases, whereas for $a>1$ it decreases. In the extremal limit corresponding to $a=2$ and $Q/M=\sqrt{2}$, the temperature vanishes, as expected.

\begin{figure}[H]
\begin{minipage}{0.49\linewidth}
\center{\includegraphics[width=0.97\linewidth]{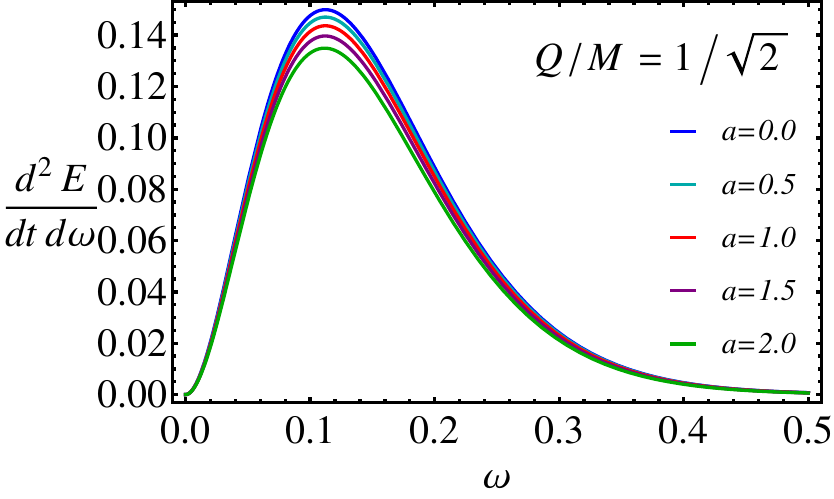}}\\ 
\end{minipage}
\hfill 
\begin{minipage}{0.50\linewidth}
\center{\includegraphics[width=0.97\linewidth]{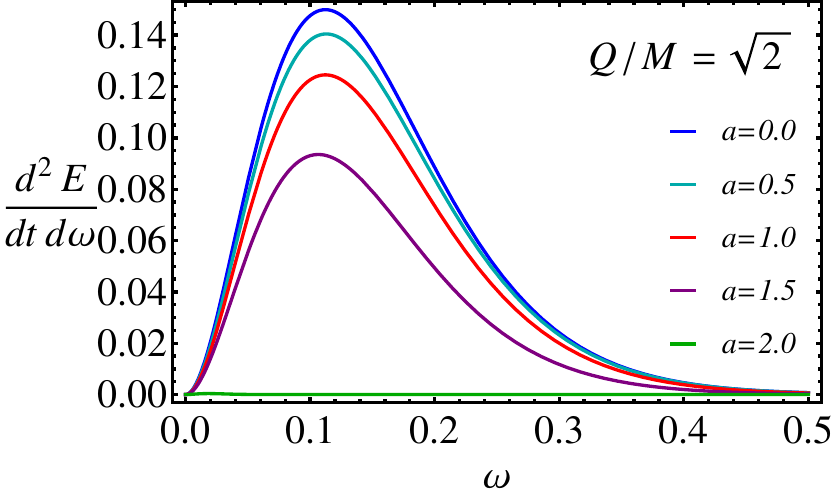}}\\ 
\end{minipage}
\caption{Emission rate of the BH  with respect to $\omega$ using $M = 1$ and $Q=1/\sqrt{2}$ (left panel) and $Q=\sqrt{2}$ (right panel) for various value of the parameter space.}
    \label{fig:emis}
\end{figure}

Examining the energy emission rate \eqref{Eq:emission} directly in terms of the parameters $a$ and $Q$ reveals the following behavior. As the dilatonic coupling $a$ increases with increasing $Q$, the overall height of the emission peak decreases, while the position of the peak frequency remains approximately unchanged (see Fig.~\ref{fig:emis}).  Moreover, in the extremal limit corresponding to $a=2$ and $Q/M=\sqrt{2}$, the vanishing temperature leads to a complete suppression of the emission rate, independent of the frequency.

\section{Strong Field Limit and Bozza Coefficients}
\label{sec:Bozz}
In the previous section, the photon sphere radius $r_{ps}$ and the corresponding shadow radius were obtained for the dilatonic black hole. These quantities determine the boundary between photons scattered to infinity and photons captured by the black hole. We now use the same critical photon orbit to analyze the deflection angle in the strong-field regime.

Following Bozza’s strong-deflection formalism \cite{bozza2002gravitational}, with
the corrected order of the remainder term \cite{tsukamoto2016strong},
the deflection angle in the limit $b\to b_{\rm cr}^{+}$ can be written as
\begin{equation}
\label{Eq:Bozza}
\alpha(b)
=
-\bar a \ln\left(\frac{b}{b_{\rm cr}}-1\right)
+\bar b
+\mathcal{O}\!\left[
\left(\frac{b}{b_{\rm cr}}-1\right)
\ln\left(\frac{b}{b_{\rm cr}}-1\right)
\right],
\end{equation}
where $b=L/E$ is the impact parameter and $b_{cr}$ is its critical value at the photon sphere.

Since the photon sphere radius has already been obtained in the previous section, we can directly evaluate the critical impact parameter. Using the impact parameter definition and substituting the dilatonic metric functions, we find
\begin{equation}
\label{Eq:b_critic}
b_{\rm cr}
=
\left.\sqrt{\frac{H(r)}{F(r)}}\right|_{r=r_{ps}} = r_{ps} \frac{\left( 1 - \frac{r_1}{r_{ps}} \right)^a}{\sqrt{1 - \frac{r_2}{r_{ps}}}}.
\end{equation}
Thus, the critical impact parameter coincides with the shadow radius measured by an observer at infinity.

The first strong-field coefficient $\bar{a}$, which controls the logarithmic divergence of the deflection angle, can be obtained from the metric functions evaluated at the photon sphere. In Bozza’s formalism it is given by
\begin{equation}
\label{Eq:a_bar_general}
\bar{a}
=
\sqrt{
\frac{2F_{ps}G_{ps}}
{H_{ps}^{\prime\prime}F_{ps}-H_{ps}F_{ps}^{\prime\prime}}
}.
\end{equation}
where the subscript $ps$ denotes evaluation at $r=r_{ps}$. Since $G(r)=1/F(r)$ for the present metric, this expression reduces after substitution to
\begin{equation}
\label{Eq:a_bar_dilatonic}
\bar{a}
=
\frac{r_{ps}-r_1}
{
\sqrt{
r_{ps}^{2}-2r_1 r_{ps}+(1-a)r_1^{2}+a r_1 r_2
}
}.
\end{equation}

This coefficient reduces to the Schwarzschild value $\bar{a}=1$ in the limit $Q=0$, where $r_1=0$, $r_2=2M$, and $r_{ps}=3M$. Therefore, deviations of $\bar{a}$ from unity reflect the influence of the dilatonic coupling and the charge on the strong-field lensing behavior.

\begin{figure}[H]
    \centering
\includegraphics[width=0.45\textwidth]{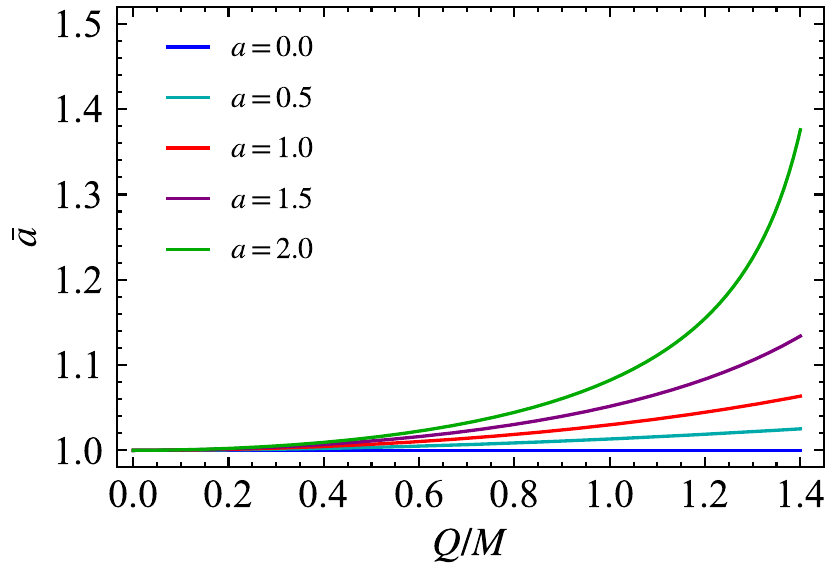}
    \vspace{0.5cm}
\caption{The Bozza strong-field coefficient $\bar{a}$ as a function of the black hole charge $Q/M$ for different values of the dilatonic coupling parameter $a$.}
    \label{fig:coef-a}
\end{figure}

To illustrate this dependence, we plot the coefficient $\bar{a}$ as a function of the black hole charge $Q/M$ for several values of the dilatonic coupling parameter $a$. The result is shown in Fig.~\ref{fig:coef-a}. One can see that for $a=0$, the coefficient remains equal to the Schwarzschild value $\bar{a}=1$ for the whole considered range of $Q/M$. For nonzero values of the dilatonic coupling, $\bar{a}$ increases with the charge. This growth becomes more pronounced as $a$ increases, indicating that the combined effect of charge and dilatonic coupling strengthens the logarithmic part of the strong-field deflection angle. In the
Reissner--Nordström limit $a=2$, the coefficient remains finite at
extremality:
$\bar a\big|_{Q/M=\sqrt{2}}=\sqrt{2}.$

Having analyzed the leading logarithmic coefficient $\bar{a}$, we now turn to the second Bozza coefficient $\bar{b}$. This coefficient describes the finite regular contribution to the deflection angle. According to Bozza’s formalism, it can be written as
\begin{equation}
\label{Eq:b_general}
\bar{b} = -\pi + b_R + \bar{a}\ln\left(\frac{2\beta_{ps}}{F_{ps}}\right)
\end{equation}
where $b_{R}$ is the regular part of the deflection integral, while $\beta_{ps}$ is the coefficient appearing in the expansion of the divergent part of the integral near the photon sphere. For the metric functions used in this work, it is given by
\begin{equation}
\beta_{ps} = \frac{H_{ps}(1-F_{ps})^2 \left( H_{ps}^{\prime\prime}F_{ps} - H_{ps}F_{ps}^{\prime\prime} \right)}{2F_{ps}^{2}H_{ps}^{\prime\,2}}
\end{equation}

The regular term $b_R$ is obtained by separating the divergent part of the deflection integral. Following Bozza’s prescription, we introduce the variable 
\begin{equation}
\label{Eq:z}
z=\frac{F(r)-F_{ps}}{1-F_{ps}}, 
\end{equation}
so that $z=0$ corresponds to the photon sphere and $z=1$ corresponds to spatial infinity. In this variable, $r$ is treated as a function of $z$, determined implicitly by
\begin{equation}
\label{Eq:zz}
F(r)=F_{ps}+(1-F_{ps})z .
\end{equation}

The regular part of the deflection integral can then be written as

\begin{equation}
\label{Eq:bR_integral}
b_R
=
\int_{0}^{1}
\left[
R(z,r_{ps})f(z,r_{ps})
-
\frac{2\bar{a}}{z}
\right]dz .
\end{equation}

\begin{equation}
\label{Eq:R_bozza}
R(z,r_{ps})
=
\frac{
2\sqrt{F(r)G(r)}
\left(1-F_{ps}\right)
\sqrt{H_{ps}}
}
{
H(r)F'(r)
}.
\end{equation}
and
\begin{equation}
\label{Eq:f_bozza}
f(z,r_{ps})
=
\left[
F_{ps}
-
F(r)\frac{H_{ps}}{H(r)}
\right]^{-1/2}.
\end{equation}

For the present metric $G(r)=1/F(r)$, and therefore $\sqrt{F(r)G(r)}=1$. The subtraction term $2\bar{a}/z$ removes the logarithmic divergence at $z=0$, so that the remaining integral is finite and can be evaluated numerically for each fixed pair of parameters ($a,Q$). In the present work, we keep $\bar{b}$ in this formal Bozza representation and focus mainly on the analytical quantities $b_{\rm cr}$ and $\bar{a}$, which determine the shadow radius and the leading logarithmic divergence of the deflection angle.

Thus, the strong-field expansion is fully characterized by the critical impact parameter $b_{\rm cr}$, the logarithmic coefficient $\bar{a}$, and the regular coefficient $\bar{b}$. The first two quantities are obtained explicitly for the dilatonic black hole, while $\bar{b}$ provides the formal finite contribution to the deflection angle.

\section{Conclusion}
\label{sec:con}
In this work, we studied the shadow and optical properties of a static dyon-like dilatonic black hole. The photon sphere radius and the corresponding shadow radius were derived and analyzed as functions of the charge $Q/M$ and the dilatonic coupling parameter $a$. It was shown that increasing these parameters decreases both radii, leading to a smaller apparent shadow.

We also compared the predicted angular diameter with the EHT data for M87$^{*}$ and Sgr~A$^{*}$. The obtained results show that current shadow measurements cannot constrain the allowed region of the dilatonic parameter space. The shadow size is sensitive to $a$ and $Q/M$, but the current data leave a wide region of the parameter space viable.

The high-frequency emission rate was estimated using the geometric-optics approximation, where the limiting absorption cross section is determined by the shadow radius. We found that increasing the dilatonic coupling reduces the peak emission intensity, while the peak position changes only weakly for the considered parameters.

Finally, the strong-field deflection regime was considered using Bozza’s formalism. An explicit expression for the leading coefficient $\bar{a}$ was obtained. This coefficient reduces to the Schwarzschild value in the corresponding limit and increases with the charge for nonzero dilatonic coupling.

These findings contribute to broader efforts to investigate strong-field gravity through black-hole shadows and gravitational lensing. Previous studies have demonstrated that EHT observations can constrain deviations from the Kerr spacetime and the parameters of alternative black-hole models \cite{2020PhRvL.125n1104P,2018NatAs...2..585M,2021PhRvD.103j4047K,2023CQGra..40p5007V}. Strong-field lensing has also been investigated in string-inspired black-hole solutions \cite{2003PhRvD..67j3009B}, while recent work has shown that naked singularities in Brans–Dicke gravity may produce shadow-like features distinct from those of Schwarzschild black holes \cite{2025EPJC...85..974P}. 

At the same time, realistic image simulations indicate that different spacetime geometries may produce similar observable features, highlighting the importance of accretion physics and observational uncertainties. In this context, our analysis provides a complementary investigation of how charge and dilatonic coupling affect the shadow, emission spectrum, and strong-field lensing properties of a dyon-like dilatonic black hole.

As a natural extension of this work, one may consider the regime $a>2$, which corresponds to 
certain modified 
dyonic black holes. Furthermore, since astrophysical black holes are expected to possess some degree of rotation, one can employ the modified Janis--Newman algorithm \cite{2014PhRvD..90f4041A,2014EPJC...74.2865A,2014PhLB..730...95A} to investigate rotating dilatonic black hole solutions in analogy with \cite{2006PhLB..639...95K, 2026JCAP...04..005H,2026JHEP...04..078B,2026arXiv260100066D}.

\subsection*{Acknowledgment}
This research is funded by the Science Committee of the Ministry of Science and Higher Education of the Republic of Kazakhstan (Grant No. AP26195301, Principal Investigator: Dr.~Manas~Khassanov).

\subsection*{Funding and/or Conflicts of interests/Competing interests}
The authors declare that they have no known competing financial, conflict and competing interests that could have appeared to influence the work reported in this paper.

\subsection*{Data Availability Statement}
Data sharing not applicable to this article as no datasets were generated or analyzed during the current study.

\bibliographystyle{spphys}
\bibliography{app.bib}
\end{document}